\documentclass[aps,prb,twocolumn,groupedaddress]{revtex4}
\usepackage{epsfig}
\usepackage{color}
\usepackage{graphicx}
\usepackage{amssymb}

\begin{document}

\title{Pairing in population imbalanced Fermion systems}

\author{M. J. Wolak$^1$, V. G. Rousseau$^2$, G. G.
Batrouni$^{1,3}$}

\affiliation{$^1$Centre for Quantum Technologies,
National University of Singapore; 2 Science Drive 3 Singapore 117542}

\affiliation{$^3$Department of Physics and Astronomy, Louisiana State
University, Baton Rouge, Louisiana 70803, USA}

\affiliation{$^3$INLN, Universit\'e de Nice-Sophia Antipolis, CNRS; 
1361 route des Lucioles, 06560 Valbonne, France}

\begin{abstract}
We use Quantum Monte Carlo (QMC) simulations to study the pairing
mechanism in a one-dimensional fermionic system governed by the Hubbard
model with attractive contact interaction and with imbalance between
the two spin populations. This is done for the uniform system and
also for the system confined in a harmonic trap to compare
with experiments on confined ultra-cold atoms. In the uniform case we
determine the phase diagram in the polarization-temperature plane and
find that the ``Fulde-Ferrell-Larkin-Ovchinnikov'' (FFLO) phase is
robust and persists to higher temperature for higher polarization. In
the confined case, we also find that the FFLO phase is stabilized
by higher polarization and that it is within the range of detection
of experiments currently underway.
\end{abstract}

\maketitle

The best understood mechanism for pair formation of fermions is the
BCS mechanism \cite{BCS} where two fermions with opposite spin and
equal but opposite momenta form a pair with zero center-of-mass
momentum. Shortly after the development of the BCS theory
of superconductivity, the question of pair formation in
polarized superconducting systems, {\it i.e.} where the populations of
the two spin states are imbalanced, was addressed independently by
Fulde and Ferrel \cite{fulde} (FF), Larkin and Ovchinnikov \cite{larkin}
(LO) and Sarma \cite{sarma}. Initially, the question was motivated
by interest in the nature of superconductivity in the presence of
a magnetic field but since then other instances where such a
mechanism intervenes have become of interest. For example, in the
interior of supermassive neutron stars, quarks of various colors may
form pairs which are not colorless thus leading to what is known as
``color superluid'' \cite{colorsf}. Another situation of major
current experimental interest is in systems of confined ultra-cold
fermionic atoms such as $^6$Li or $^{40}$K. Such experiments have
now reported the presence of pairing in the case of unequal
populations \cite{zwierlein06,partridge06} in three-dimensional
cigar shaped traps and in one dimensional traps \cite{hulet}.
However, the precise nature of the pairing has not yet been
elucidated experimentally.

On the theoretical side, many methods have been used ranging from mean
field \cite{mft} to effective Lagrangian \cite{son06} to Bethe ansatz
\cite{orso}. The two competing mechanisms for pair formation in the
population imbalanced case are the FFLO and the Sarma mechanisms. In
the former, the bosonic pairs form with non-zero center of mass
momentum equal to the difference in the Fermi momenta,
$|k_{F1}-k_{F2}|$, of the two populations. This leads to the formation
of a standing wave in the order parameter whose wave vector is
$|k_{F1}-k_{F2}|$. With the Sarma mechanism, majority fermions whose
momenta equal the Fermi momentum of the minority, are promoted to
higher momentum levels thus forming a breach in the Fermi distribution
of the majority population. This breach allows majority fermions with
momentum equal to the Fermi momentum of the minority to pair up with
minority fermions forming pairs with zero center of mass
momentum. Extensive numerical work on the one dimensional system
using Quantum Monte Carlo (QMC) \cite{batrouni08,casula}, and the
Density Matrix Renormalization Group (DMRG) \cite{dmrg}) has
demonstrated that, in the ground state, population imbalance leads to
a robust FFLO phase over a very wide range of polarization and
interaction strengths. The Sarma phase was not detected in these
numerical works.

The stability of the FFLO phase at finite
temperatures has been adressed with mean field calculations
\cite{mftfiniteT} which can be unreliable in low dimension where
quantum fluctuations are large. Recently we addressed this question
\cite{ggb2010} using exact quantum Monte Carlo (QMC) simulation which
is the focus of this presentation.

In order to study the pairing mechanism of fermions in an optical
lattice, we consider the one-dimensional fermionic Hubbard
Hamiltonian,
\begin{eqnarray}
\label{Hamiltonian}
H&=&-t\sum_{i\, \sigma} (c_{i\,\sigma}^{\dagger}
c_{i+1\,\sigma}^{\phantom\dagger} + c_{i+1\,\sigma}^\dagger c_{i \,
\sigma}^{\phantom\dagger}) - \sum_i ( \mu_1\hat n_{i\,1} +
\mu_2 \hat n_{i\,2}) \nonumber \\ 
\nonumber
&&+U \sum_{i} \left(\hat n_{i\,1}-\frac{1}{2}\right)\left(\hat
n_{i\,2}-\frac{1}{2}\right)\\ 
&&+V_T \sum_{i} \left(x_i-\frac{L}{2}\right)^2 \left(\hat n_{i \,1} +
\hat n_{i \, 2}\right) 
\end{eqnarray}
where $c_{i\,\sigma}^\dagger$ and $c_{i\, \sigma}^{\phantom\dagger}$
are fermion creation and annihilation operators on lattice site $i$
satisfying the usual anticommutation relation,
$\{c^{\phantom\dagger}_{i\,\sigma},c^\dagger_{j\,
\sigma^{\prime}}\}=\delta_{i,j}\delta_{\sigma,\sigma^\prime}$. The
fermionic species are labeled by $\sigma=1,2$ and $\hat
n_{i\,\sigma}=c_{i\,\sigma}^\dagger c_{i\, \sigma}^{\phantom\dagger}$
is the corresponding number operator. The energy scale is set by
taking the hopping parameter $t=1$.  The contact interaction strength
$U$ is negative since we are interested in pair formation in the
attractive model. The last term describes the confining harmonic trap
which is centered at the midpoint, $L/2$, of the $L$-site lattice.  We
take periodic boundary conditions. Our QMC results were obtained using
the Determinant QMC algorithm \cite{DQMC} (DQMC) and the Stochastic
Green Function (SGF) technique \cite{SGF}. Specifically, we used the
DQMC algorithm to determine the phase diagram in the
polarization-temperature ($P,T$) plane because of its good convergence
properties for large systems. This algorithm functions in the grand
canonical ensemble where the chemical potential for each species is
tuned to obtain the desired polarization. When the populations are
imbalanced, this algorithm suffers from the sign problem even with the
attractive interaction, as is the case here. However, in our
simulations, the average sign never went below around $0.3$ at the
highest polarizations and lowest temperatures. This allowed us to
study the system under rather extreme conditions. The SGF algorithm
was used for all other simulations including all simulations of the
confined system because this algorithm functions in the canonical
ensemble, where the populations are fixed, which corresponds to the
experimental situation. In addition, we have verified \cite{mass} that
for the system sizes and fillings we simulated, the grand canonical
(DQMC) and canonical (SGF) ensembles give the same results. Typical
simulations, with DQMC or SGF, took from five to seven days each on a
$3$GHz processor. The simulations were performed locally on our
cluster with $102$ cores.

\begin{figure}[h]
\begin{center}
\includegraphics[width=6cm,clip]{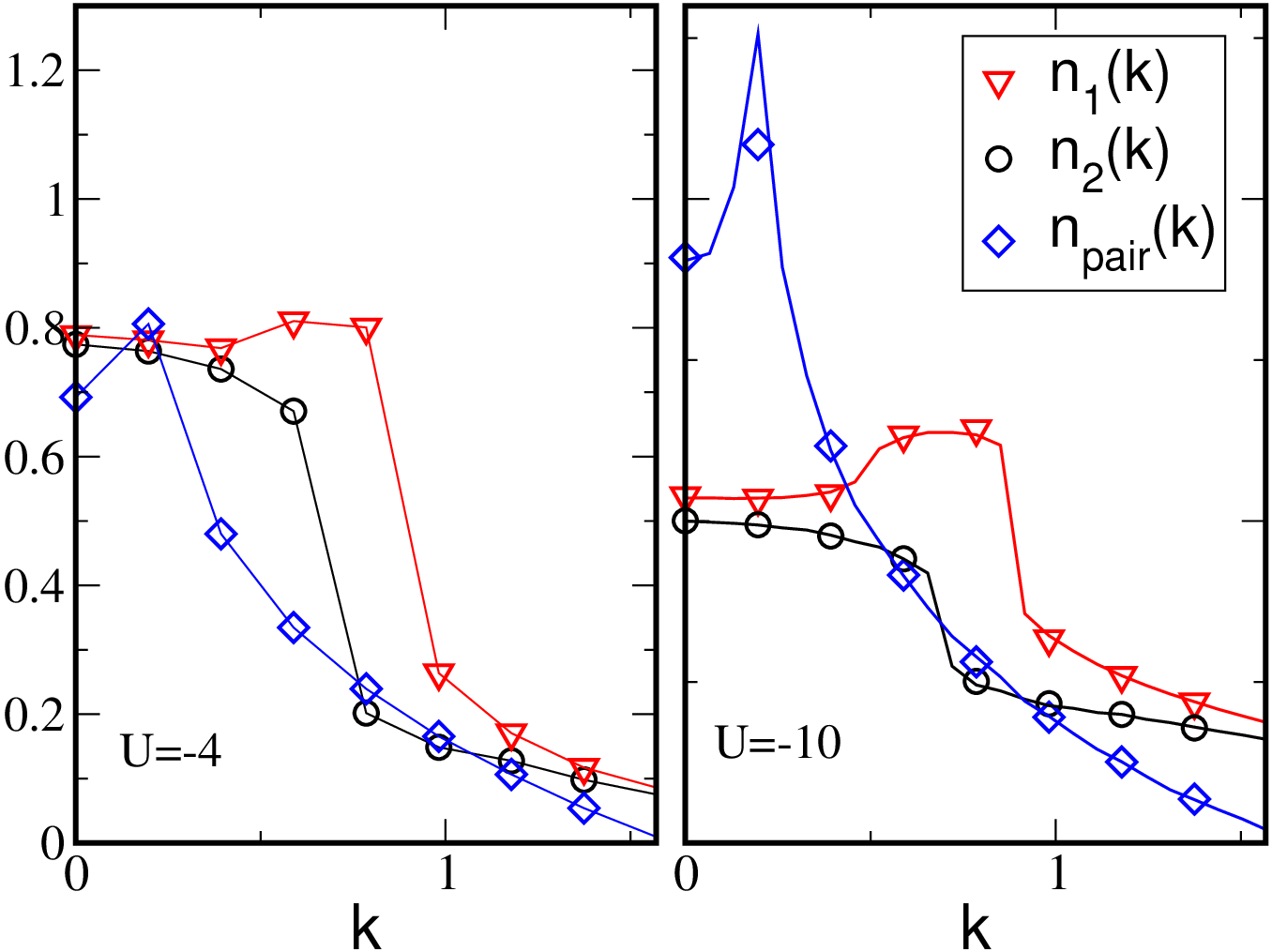}
\includegraphics[width=6cm,clip]
{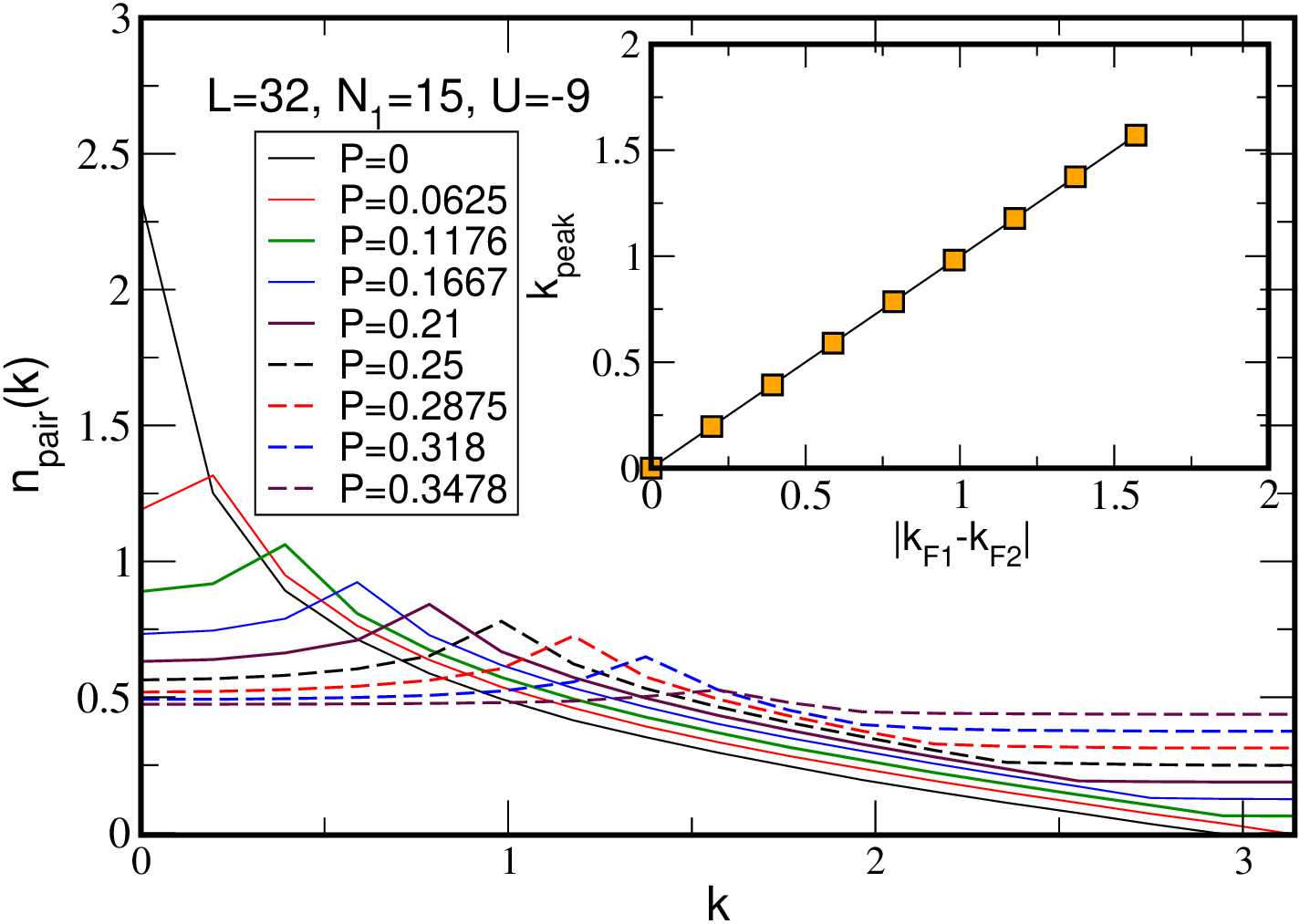}
\end{center}
\caption{Top: The momentum distributions for the majority, minority and
pair populations in a system of $L=32$ sites, majority population
$N_1=9$, minority $N_2=7$, $\beta=64$. The pair momentum distribution
exhibits a distinct maximum at nonzero momentum. Bottom: The pair
momentum distribution for several polarizations $P$. The inset shows
that the position of the FFLO peak is equal to 
$|k_{\rm F1}-k_{\rm F2}|$.}
\label{uniform}
\end{figure}

Using these algorithms, we calculate the real space Green
functions of each species, $G_\sigma$, and the pair green function,
$G_{\rm pair}$,  
\begin{eqnarray}
\label{gfct}
G_\sigma(l) &=& \langle c_{j+l\,\sigma}^\dagger c_{j
\,\sigma}^{\phantom\dagger} \rangle,\\
\label{pairgfct}
G_{\rm pair}(l) &=& \langle
\Delta^{\dagger}_{j+l}\,\Delta_{j}^{\phantom\dagger} \rangle,\\ 
\label{pairoperator}
\Delta_j &=& c_{j \, 2} \,c_{j \, 1},
\end{eqnarray}
where $\Delta_j$ destroys a pair on site $j$. The Fourier transform of
$G_{\sigma}(l)$ yields the momentum distributions $n_{\sigma}(k)$ and
the transform of $G_{\rm pair}(l)$ leads to the pair momentum
distribution, $n_{\rm pair}(k)$, a central quantity in this work. In
the non-interacting limit, the Fermi momentum of a population is given
by $k_{F\sigma}=\frac{N_{\sigma}-1}{2}\frac{2\pi}{L}$, where $L$ is
the number of sites and $N_{\sigma}$ the number of particles. The
Fermi energy is given in the uniform case by $\epsilon_F=tk_F^2=T_F$
where $T_F$ is the Fermi temperature. The polarization is defined
as $P=(N_1-N_2)/N$ where $N=N_1+N_2$ is the total number of particles.

In Fig.\ref{uniform} we show results for the uniform system
\cite{batrouni08}, $V_T=0$. The top panels show the momentum
distributions for the majority and minority populations and for the
pairs at $U=-4t,\, -10t$. We see that the Fermi ``surfaces'' for
the two populations are sharply defined and that the peak in the pair
momentum distributions is at nonzero momentum, $k_{\rm peak}$: This is
a clear signal for FFLO pairing. It is interesting to note the
deformation of the momentum distribution of the majority population: A
bump develops in $n_1(k)$ for $k>k_{F2}$. As the Fermi distributions of
the minority and majority start to match up more closely as $U$ becomes
more attractive, the excess unpaired majority particles populate the
states with momenta larger than the minority Fermi momentum. This
give the bump in $n_1(k>K_{F2})$. The lower
panel of Fig.\ref{uniform} shows $n_{\rm pair}(k)$ for several
polarizations. It is clearly seen that as $P$ increases so does $k_{\rm
peak}$, the momentum at which $n_{\rm pair}(k)$ peaks. In the inset we
display $k_{\rm peak}$ versus $|k_{\rm F1}-k_{\rm F2}|$ which shows
that in fact $k_{\rm peak} = |k_{\rm F1}-k_{\rm F2}|$ as predicted by
the FFLO scenario. The nonzero value of the pair center of mass
momentum means that its Fourier transform $G_{\rm pair}(l)$ (Eq.
\ref{pairgfct})) oscillates \cite{batrouni08} with wavelength
$\lambda =2\pi/|k_{\rm peak}|$. This means that in the FFLO phase  the
system is not homogeneous, it consists of pair-rich regions and
regions depleted in pairs but rich in the excess unpaired majority
population.

\begin{figure}
\begin{center}
\includegraphics[width=8cm,clip]{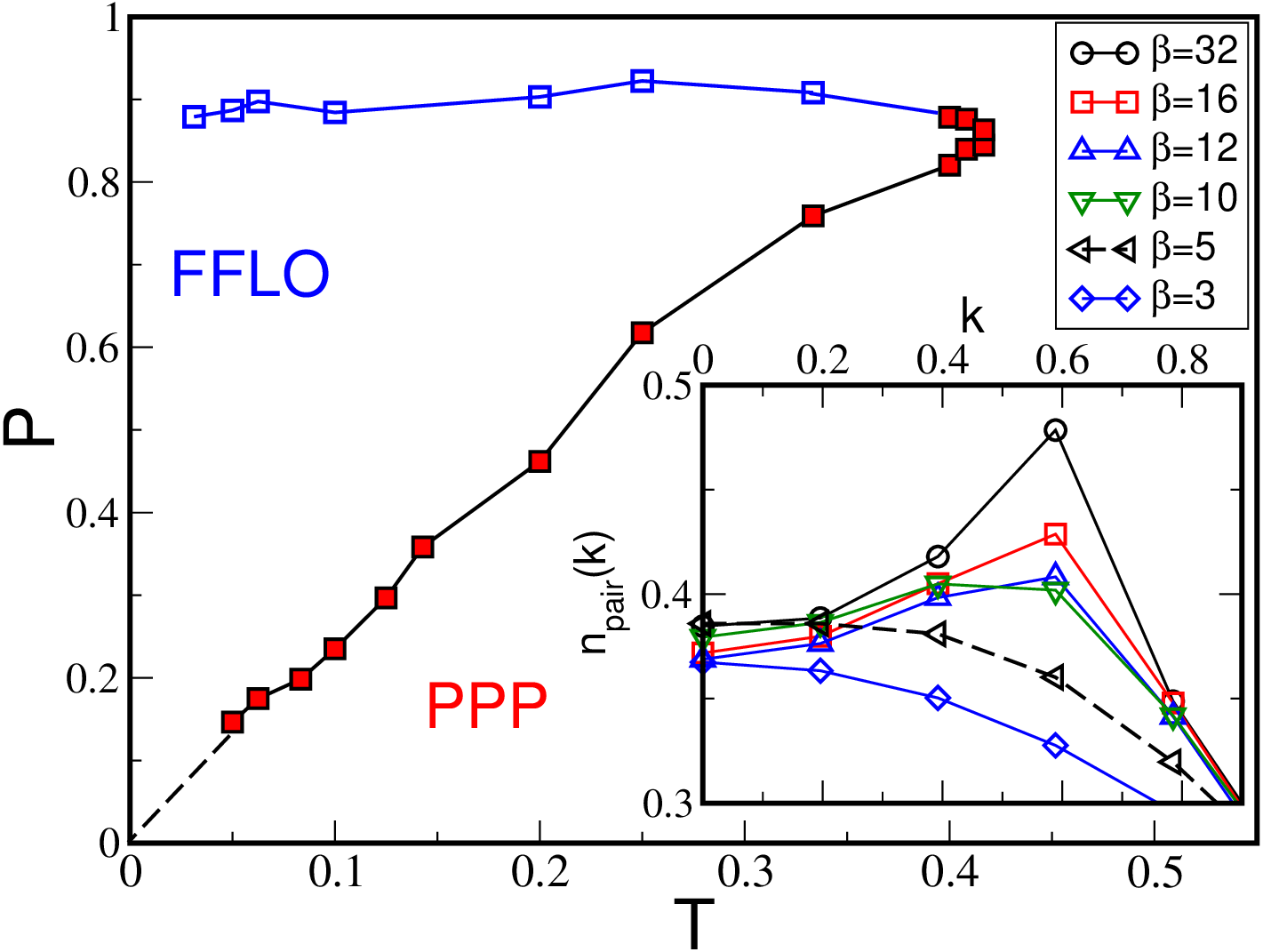}
\caption{The phase diagram at quarter filling, $N=L/2$, with $U=-3.5t$
and $L=50$. The Fermi temperature at this filling is $T_{\rm F}=0.5t$.
We see that the FFLO phase is robust and persists over a wide range of
$P$ and $T$. Increasing $P$ stabilizes FFLO up to higher $T$. PPP
stands for Polarized Paired Phase, see text for a discussion. The inset
shows the behavior of the FFLO peak as $T$ is increased for the case
$N_1=13$, $N_2=7$, {\it i.e.} $P=0.3$. We see that the FFLO peak goes
down as $T$ increases and vanishes at $\beta =1/T\approx 5$.}
\label{phasediag}
\end{center}
\end{figure}

Such calculations were done for several values of $U$ using DMRG
\cite{dmrg} and QMC \cite{batrouni08} and lead to the conclusion
that, in the ground state, the polarized system is always in the FFLO
phase and the Sarma mechanism does not intervene. The question then
arises as to the effect of finite temperature: Will FFLO survive at
$T\neq0$ and how robust is this phase?

To this end, we studied, using QMC simulations, what happens to the
FFLO peak as the temperature is increased. The inset in
Fig. \ref{phasediag} shows this peak for the case $N_1=13$ and $N_2=7$
($P=0.3$). We see that as $T$ increases, the FFLO peak gets lower and
eventually disappears at $\beta=1/T\approx 5$ which signals
the destruction of the FFLO phase by thermal fluctuations. But how is
FFLO destroyed? The two possibilities are (a) the pairs are broken or
(b) the pairs are still formed but thermal fluctuations make the system
homogeneous resulting in a peak at $k=0$ for $n_{\rm peak}(k)$. To
destroy the pairs, the thermal energy should be of the order of the pair
binding energy, {\it i.e.}  $T\sim |U|$. However, we see from Fig.
\ref{phasediag} that FFLO is always destroyed at $T\ll |U|$.
We conclude, therefore, that thermal fluctuations destroy FFLO by making
the system homogeneous not by destroying the pairs; we denote this
phase by Polarized Paired Phase (PPP). In this way, we determine the
phase diagram of the system, shown in Fig.\ref{phasediag} at quarter
filling $N=N_1+N_2=L/2$. The figure shows clearly that the FFLO phase
is very robust extending over a wide range of $P$ and $T$. Increasing
$P$ stabilizes FFLO up to higher $T$ while at low $P$ even a small
increase in $T$ destroys it. Consequently, for small $P$ one needs to
simulate the system at exceedingly low $T$ to see FFLO. This increases
the simulation time and sets a limit on the lowest practical $P$. The
dotted line in Fig. \ref{phasediag} schematizes the phase boundary at
very low $P$. At the other extreme, $P\to 1$, the system is primarily
made of one population with very few minority particles. This makes the
FFLO signal very difficult to see. The open symbols in Fig.
\ref{phasediag} denote the highest $P$ we were able to examine, the
system up to those values is still FFLO.

\begin{figure}
\begin{center}
\includegraphics[width=8cm,clip]
{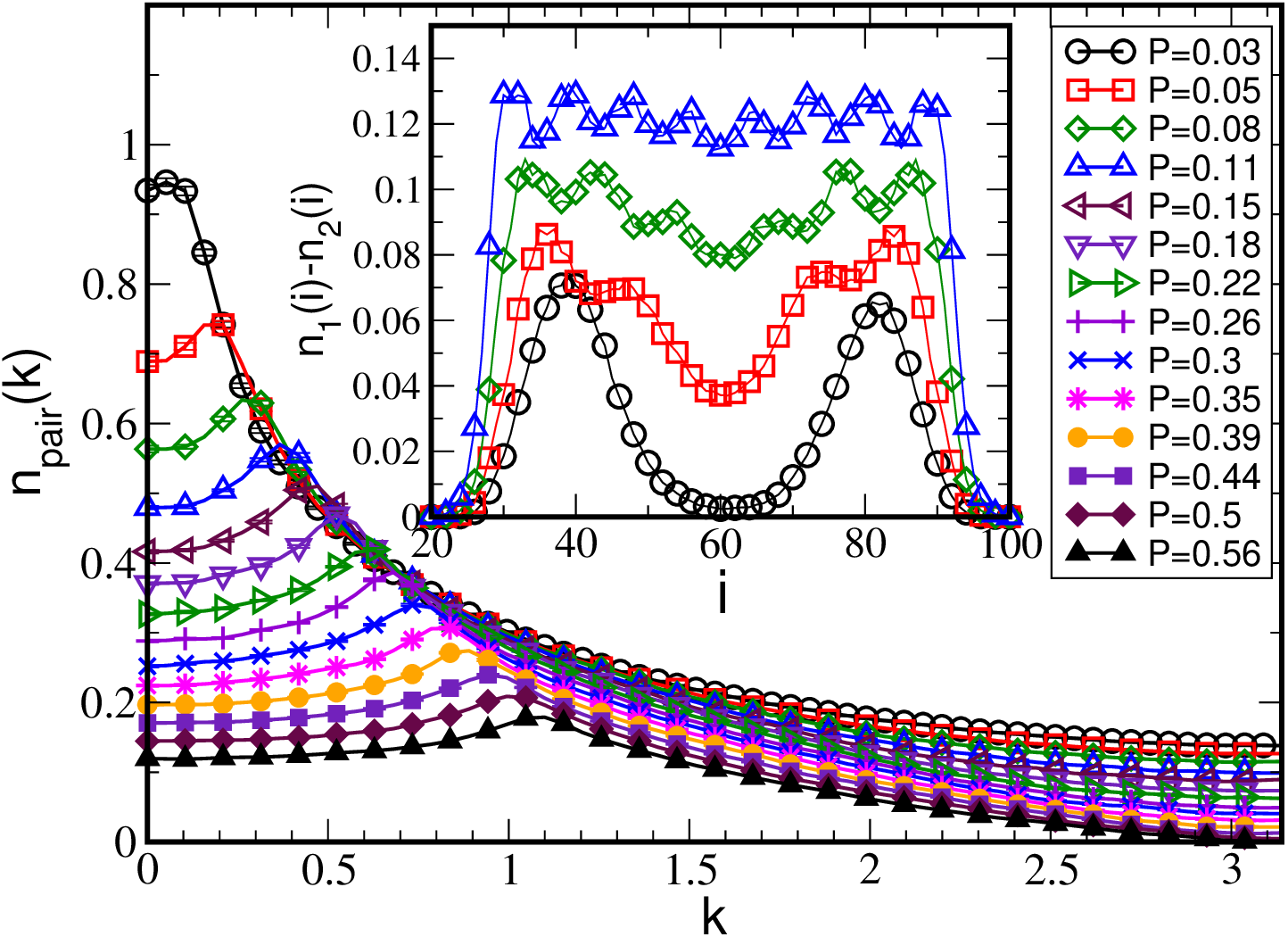}
\caption{The pair momentum, $n_{\rm pair}(k)$, versus $k$ for several
polarizations of the confined system, $V_T=0.0007t$, $L=120$, $U=-10t$
and $\beta=64$. The majority population is $N_1=39$ and the minority,
$N_2$, is varied to tune $P$. Inset: Profile of the populations
difference for the smallest four polarizations. See the text for a
discussion of the oscillations.}
\label{trappairmom}
\end{center}
\end{figure}

Continuing earlier work in higher dimension
\cite{zwierlein06,partridge06}, the Rice group \cite{hulet}
recently reported on experiments in arrays of one dimensional tubes
of confined Fermionic atoms ($^6$Li) with imbalanced
populations. Along the tube, the atoms were confined with a
trap frequency $\omega_z=2\pi\times 200$Hz; in the central tube,
the total number of atoms at zero polarization was approximately $250$
and the temperature was estimated at $T/T_F\approx 0.1$ where the Fermi
temperature $T_F$ is obtained from the Fermi energy
$\epsilon_F=(1/2+\hbar \omega)N/2$. The pair
binding energy, $\epsilon=\hbar^2/ma_{1D}$ (where $a_{1D}$ is
the effective one-dimensional scattering length), was estimated to be
$\epsilon/\epsilon_F\approx 5.3 $.

We now present QMC results for the fermionic Hubbard
model, Eq.(\ref{Hamiltonian}), in the presence of the confining trap.
To make contact with the experiment we introduce
the trapping potential in Eq.(\ref{Hamiltonian})
$V_T=0.0007t$ which corresponds to $\hbar\omega_z=2\sqrt{tV_T}$. The
total number of particles in our simulations for balanced populations
is $78$, to be compared with $250$ in the experiment. We performed our
simulations in the temperature range $0.008\leq T/T_F\leq 0.25$ which
includes the temperature at which the experiments were performed,
$T/T_F=0.1$. In addition, to place our system in the same coupling
parameter regime as the experiments, we present our results for a
coupling strength of $U=-10t$. $U$ is the ``pair binding energy'' and
the value we have chosen gives $|U|/\epsilon_F=4.8$, close to the
experimental value.

In Fig.\ref{trappairmom} we show, as we did in the uniform case
Fig.\ref{uniform}, the pair momentum distribution, $n_{\rm pair}(k)$
for several $P$ values. We see that $k_{\rm peak}\neq 0$ and that
its value increases with $P$ as predicted by the FFLO picture.
These results were obtained at $\beta=64$ and essentially represent
the ground state behavior of the system. Therefore, the presence of
the trap does not change the nature of the phase in the ground state,
it remains FFLO when the system is polarized and is robust. The inset
in Fig. \ref{trappairmom} shows the difference between the majority
and minority density profiles for the four lowest $P$ values we
examined. We see that the difference is oscillatory; the wavelength
of the oscillations is in fact given by $\lambda=2\pi/|k_{\rm
F1}-k_{\rm F2|}|$. For example, we see that for the smallest $P$ we
show, one wavelength fits in the system and for the $P=0.11$ case
four wavelengths fit. This is a nice visual confirmation that the
FFLO state is not homogeneous.

\begin{figure}[h]
\begin{center}
\includegraphics[height=6cm,width=4cm,clip]
{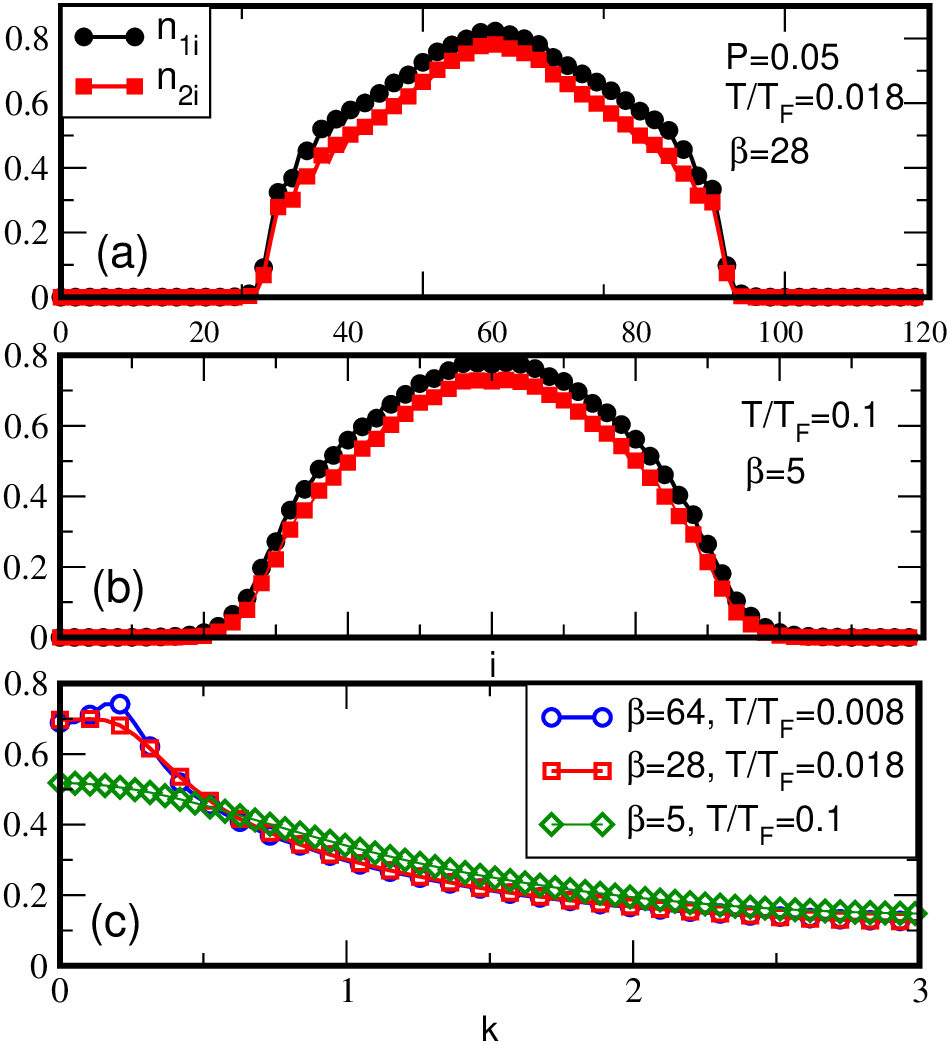}
\hskip -0.1cm \includegraphics[height=6cm,width=4cm,,clip]
{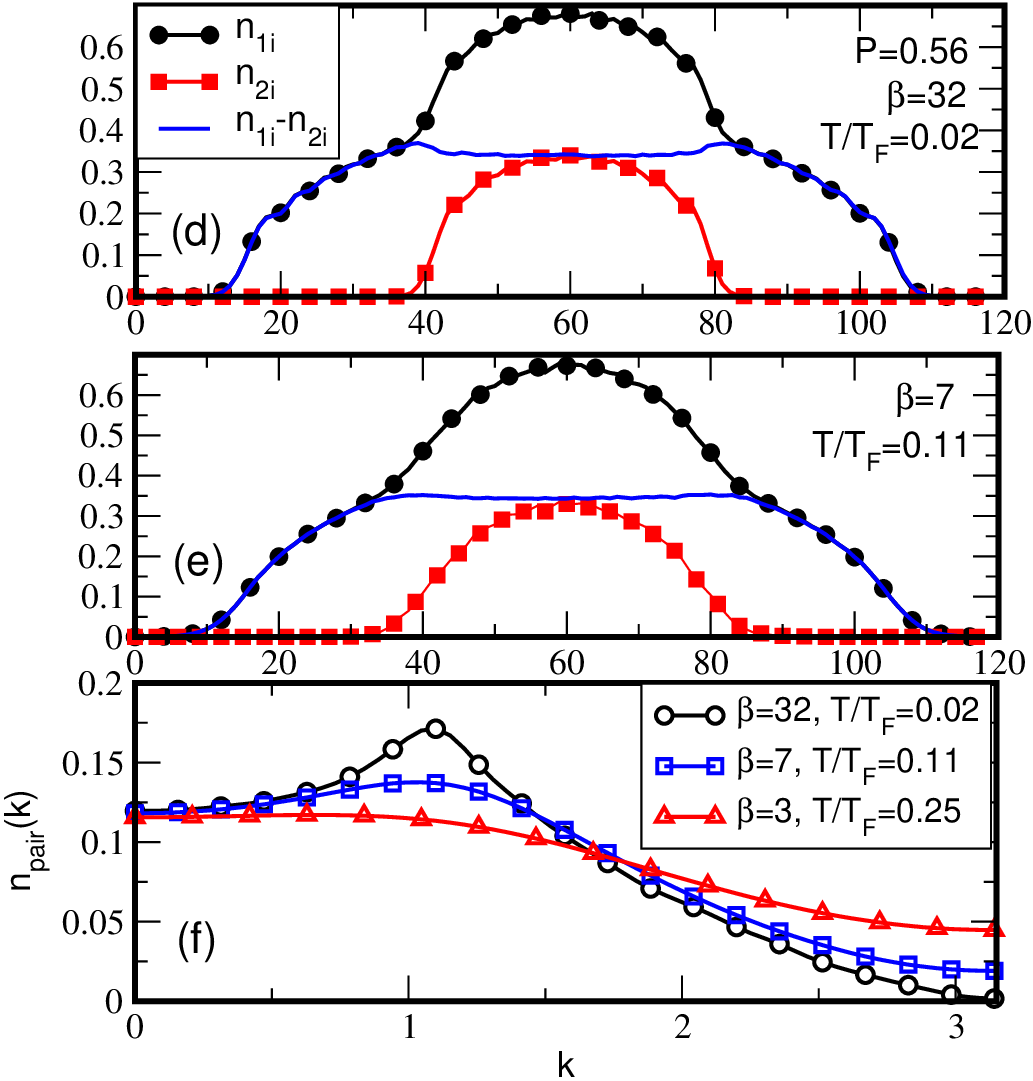}
\caption{Density profiles for $P=0.05$ at $T/T_{F}=0.01$ (a) and
$T/T_F=0.1$ (b). The pair momentum distribution for three $T$ values
(c) shows that the FFLO peaks disappears for $T/T_F=0.018$ which is
much lower than what is feasible experimentally. The high polarization
case, $P=0.56$ is shown in (d) and (e) for $T/T_F=0.02$ and $0.11$.
The pair momentum distribution (f) shows that for $P=0.56$ the FFLO
peak survives even for $T/T_F>0.1$, the experimental value.}
\label{trapprofiles}
\end{center}
\end{figure}

The effect of finite temperature is examined, as before, by studying
the behavior of the FFLO peak as $\beta=1/T$ is decreased
(increased). We show in Fig.\ref{trapprofiles} two $P$ cases. On the
left, (a) and (b) show how the density profiles, {\it i.e.} the
local density in the trap, for $P=0.05$ change as the system is heated.
We see that the profiles get more rounded as $T$ increases. The
pair momentum
distribution (c) shows that the FFLO peak disappears by the time
$T=0.018T_F$. This temperature is very low and is not accessible
experimentally. However, the high polarization case, $P=0.56$, shown in
(d) and (e) behaves differently. Here too, the profiles get rounded as
$T$ increases, but we see in (f) that the FFLO peak survives for
$T>0.1T_F$, the experimental value. This means that with presently
attainable experimental temperatures, the FFLO phase may be observed.
This increased stability of FFLO with increased $P$ is consistent with
the phase diagram we found for the uniform case, Fig.\ref{phasediag}.

In this paper we have examined the pairing mechanism in fermionic
systems with imbalanced populations both in the absence and presence of
a confining trap which breaks translational invariance. We showed that
the dominant pairing mechanism in the ground state and also at finite
temperatures is FFLO where the pairs form with a nonzero center of mass
momentum. This is revealed clearly by a peak at nonzero momentum, $k_{
\rm peak}=|k_{\rm F1}-k_{\rm F2}|$, in the pair momentum distribution.
The behavior of this peak is studied as a function of the temperature
and also the polarization, $P$. We showed that increasing $P$ stabilizes
FFLO up to higher temperatures and, in the confined case relevant to
experiments on ultra-cold fermionic atoms, places this phase within
reach.

Acknowledgments
This work was supported by: an ARO Award W911NF0710576 with funds from
the DARPA OLE Program; by the CNRS-UC Davis EPOCAL joint research
grant; by NSF grant OISE-0952300; by the France-Singapore Merlion
program (PHC Egide, SpinCold 2.02.07 and FermiCold 2.01.09) and the
CNRS PICS 4159 (France). Centre for Quantum Technologies is a Research
Centre of Excellence funded by the Ministry of Education and the
National Research Foundation of Singapore.

\end{document}